\documentclass[aps,pre,onecolumn,superscriptaddress]{revtex4}

\begin{document}

\title{Low temperature amorphous solids: mean field theory and beyond}\label{ch1}

\author{Pierfrancesco Urbani}
\affiliation{Universit\'e Paris-Saclay, CNRS, CEA, Institut de physique th\'eorique, 91191, Gif-sur-Yvette, France}

\begin{abstract}{Amorphous solids at low temperature display unusual features which have been escaped a clear and unified comprehension. 
In recent years a mean field theory of amorphous solids 
constructed in the limit of infinite spatial dimensions has been proposed.
I will briefly review what is the outcome of this theory focusing the low temperature phase and discuss some perspectives to go beyond the mean field limit.}
\end{abstract}

\maketitle

\section{Introduction}
Amorphous solids at low temperature display unusual features which have been escaped a clear and unified comprehension
since the early seventies when they have been first identified in experiments \cite{zeller1971thermal}. 
In recent years a mean field theory of amorphous solids 
constructed in the limit of infinite spatial dimensions has been proposed.
I will review what is the outcome of this theory focusing on a specific result:
depending on the nature of the 
microscopic degrees of freedom, their interaction potential and system preparation, 
amorphous solids may undergo to a so-called Gardner transition under a set of 
physical conditions (low temperature, high pressure, sufficiently high strain deformation).
At this point stable glasses become marginally stable, in the sense that they are very sensible
to external perturbations.

We will review the mean field theory of the Gardner transition and the nature of the Gardner phase, its connection
with spin glasses, and discuss where we stand on the quest for an extension
of these results beyond the mean field limit.
We will mainly focus on classical sytems without discussing the effect of quantum fluctuations.
We will also restrict the attention to the theoretical picture emerging from the infinite dimensional limit with very little discussion on how
this compares with simulations, experiments and other approaches.
The interested reader can look at other chapters in this volume and recent reviews in the literature.

\section{Theory of glasses in infinite dimensions}\label{sec1.1}
The equilibrium thermodynamic route for rigidity requires symmetry breaking. 
Cold simple liquids turn into solid crystals when translational symmetry is broken.
However, in fact, many solids in nature are amorphous. At the microscopic level
the degrees of freedom are arranged in a disordered manner.
Glasses are one of the main examples. In this case the route for rigidity goes through an out-of-equilibrium pathway.
When cooled down, complex liquids avoid crystallization
and get stuck in a metastable branch of their equation of state where they become viscous supercooled liquids.
Cooling down further these systems, the viscosity increases of several orders of magnitude in a short interval
of temperatures. 
At some point they become so viscous that are effectively rigid and form a glass \cite{lubchenko2007theory,cavagna2009supercooled, wolynes2012structural, leuzzi2007thermodynamics}.

The theoretical description of the glass transition has been very debated and has triggered several different approaches. 
In the late eighties, in a series of groundbreaking works, it was proposed by T. Kirkpatrick, D. Thirumalai and P. Wolynes, 
\cite{kirkpatrick1987p, kirkpatrick1987stable, kirkpatrick1987connections, kirkpatrick1987dynamics, kirkpatrick1989scaling} that 
the glass transition was driven by the appearance of an exponential number of metastable glassy states.
This picture was built on the fundamental observation that the solution of a class spin glass models exhibiting a dynamical one-step-replica symmetry breaking (1RSB)
transition pathway (notable examples are Potts spin glasses, $p$-spin glasses with $p>2$) had essentially the same phenomenology of
structural glasses. This connection was completely non-obvious and had extremely far reaching consequences giving birth to the Random First Order Transition theory (RFOT) of the glass transition.

It took a series of fundamental works \cite{franz1995recipes,cardenas1998glass,cardenas1999constrained,monasson1995structural,mezard1999thermodynamics}, mainly devoted to understand how to adapt the replica method to systems without quenched disorder,
to show that the scenario for the glass transition proposed at the beginning
was exactly realized in the limit of infinite spatial dimensions \cite{kurchan2012exact, parisi2020theory}.

A detailed review of the solution of simple glass models in infinite spatial dimensions can be found in recent works \cite{parisi2020theory, charbonneau2017glass}.
Here I will describe the main essential steps of the construction.

The simplest approach is to start from the Franz-Parisi construction \cite{franz1995recipes, barrat1997temperature},
but alternative ways exist \cite{monasson1995structural}. 

We consider a system of $N$ interacting degrees of freedom in $d$-dimensions. The degrees of freedom are the positions of
particles and are denoted by $\underline x_i=\{x_{i1}, \ldots, x_{id}\}$ with $i=1,\ldots,N$. 
They interact through a pairwise interaction potential so that the total Hamiltonian of the system is
\begin{equation}
H[X]=\sum_{i<j}^N v(|\underline x_i-\underline x_j|)     \ \ \ \ \ X=\{\underline x_i\}_{i=1,\ldots, N}\:.
\end{equation}
We introduce the notation $\hat H[X,\underline c]$ to denote the generalized Hamiltonian with its dependence on the control parameters $\underline c$. For example, if the system is in equilibrium at an inverse temperature $\beta = 1/T$ and density $\rho$ we will denote it by $\hat H= \beta H$ and $\underline c =\{\beta, \rho\}$.
The equilibrium partition function is therefore $Z[\underline c]= \int dX \exp\left[-\hat H[X,\underline c]\right]$.

As a first approximation, a glass is an arrested
phase where particles are caged by neighbors and cannot diffuse. Defining the mean square displacement between two configurations
in phase space as
\begin{equation}
\Delta[X,Y] = \frac{1}{N} \sum_{i=1}^N |\underline x_i - \underline y_i|^2
\end{equation}
one can consider the following large deviation function, the so-called Franz-Parisi potential, defined as
\begin{equation}
V_{FP}[\hat \Delta, \underline c_m, \underline c_s] = -\int  dY\frac{e^{-\hat H[Y,\underline c_m]}}{N Z[\underline c_m]} \ln \int dX e^{-\hat H[X,\underline c_s]}\delta(\hat \Delta -\Delta(X,Y))\:.
\end{equation}
It is useful to call the $Y$-system as the master or parent glass, while the $X$-system is the slave system. Intuitively, the Franz-Parisi potential looks at the tendency of the slave system to be closer to the master one. Practically, the master system acts as quenched disorder for the slave one.

We first consider the case where $\underline c_s=\underline c_m$. To fix the ideas we will focus on hard spheres where  the only control parameter is the packing fraction $\phi$, namely the fraction of space occupied by the spheres. The corresponding interaction potential is formally infinite as soon as spheres overlap and zero otherwise.
We underline that all what follows can be repeated for thermal systems by replacing $\phi$ with the temperature $T$. 

We are interested in looking at the profile of $V_{FP}$ as a function of $\underline c_m = \phi$. 
We can identify three regimes. For sufficiently small packing fraction $\phi<\phi_d$ the Franz-Parisi potential has a global minimum at $\hat \Delta\to \infty$. This means that the system is an ergodic liquid because there is no pinning effect of the master system on the slave one. When $\phi\in[\phi_d,\phi_K]$ the Franz-Parisi potential develops a metastable minimum at a finite value of $\hat \Delta=\hat \Delta_r<\infty$. This is the dynamical glass phase and the point $\phi_d$ corresponds to the dynamical/mode-coupling glass transition. In the regime $\phi\in[\phi_d,\phi_K]$, $V_{FP}$ has a local minimum for finite $\hat \Delta$ which means that particles in the system are caged. When $\phi=\phi_K$ the local minimum of the Franz-Parisi potential becomes the global one and the system is in the ideal glass state so that $\phi_K$ corresponds to the Kauzmann transition. This dynamical 1RSB phase found in hard spheres coincides with the one of simple mean field spin glass models as forseen by Kirkpatrick, Thirumalai and Wolynes and thereofore provides a direct evidence that in infinite dimensions the glass transition is realized within the RFOT scenario.

This construction can be generalized to $\underline c_m\neq c_s$. When $\underline c_m$ is such that the master system is in a dynamical 1RSB region, this corresponds to selecting an equilibrium glass state and looking at what happens when it is perturbed. In infinite dimension one can get access to physical perturbation schemes, like cooling, compression, shear deformation. This way, the theory in infinite dimension becomes a tool to understand the behavior of glassy states when, as a firts approximation, activation is neglected. This formalism has been pushed forward to obtain a complete theory in $d\to \infty$ of simple colloidal glasses (hard spheres) under compression \cite{rainone2015following,rainone2016following}, or simple thermal glasses under cooling \cite{biroli2018liu,scalliet2019marginally} and strain deformation \cite{biroli2016breakdown, urbani2017shear,biroli2018liu,altieri2018microscopic, altieri2019mean}. This has allowed to obtain a first principle approach to study low temperature glasses or high pressure colloidal glasses close to the jamming point, and sheared amorphous solids and the corresponding yielding transition.

The computation of the Franz-Parisi potential can be performed in full details in infinite dimensions through the replica method, see \cite{parisi2020theory}. The $d\to \infty$ limit allows to reduce it to a saddle point computation for a set of order parameters. In addition to $\hat \Delta$ that minimizes $V_{FP}$ one needs to introduce a mean square displacement matrix $\Delta_{ab}$ which encodes for the mean square displacement of $n\to 0$ different replicas of the slave system all close to the same configuration of the master one.

This mean square displacement matrix is a generalization of the overlap matrix in spin glasses \cite{sherrington1975solvable, mezard1987spin} to particle systems and it represents a measure of similarity between different configurations subjected to the very same quenched disorder.

We will focus on what happens when the parent, master, glass has control parameters such that it is in the dynamical 1RSB phase. 
Again we will consider what happens for hard spheres under compression but the same can be repeated in thermal systems by decreasing the temperature.

We will consider a parent glass at a packing fraction $\phi_m\in[\phi_d,\phi_K]$ and compress it by considering a slave system at $\phi\geq \phi_m$. This can be realized for example by increasing the diameters of the spheres of the slave system as compared to the master one.
When $\phi=\phi_m$, the solution of the saddle point equations, are found within a replica symmetric ansatz. In this case one can parametrize the off-diagonal part of the overlap matrix by a constant $\Delta_{a\neq b}=\Delta_{EA}$ whose saddle point value coincides with the saddle point one for $\hat \Delta = \Delta_{r}=\Delta_{EA}$.

We can now consider what happens when the system is compressed. 
Depending on the parent glass packing fraction $\phi_m$, there exist a packing fraction $\phi_G(\phi_m)$, such that the replica symmetric assumption is correct in the interval $\phi\in[\phi_m,\phi_G(\phi_m)]$.
Therefore in this case the glass state can be thought as an ergodic metabasin: restricted equilibration within this portion of phase space takes finite time 
and correlations among degrees of freedom are short ranged\footnote{Note that, as usual in infinite dimensions, instead of correlation functions one can study integrated response functions or susceptibilities. In the stable glass phase the system responds as an amorphous elastic medium.}. This is the stable glass phase.

However, increasing the pressure above $\phi_G(\phi_m)$, the replica symmetric assumption is no longer correct and one has replica symmetry breaking within a glass basin.
This point is the Gardner transition.

\section{The Gardner transition}
The Gardner transition marks the point where a stable glass enters in a phase where replica symmetry is broken. 
This transition was first found in the context of mean field spin glasses in \cite{gardner1985spin, gross1985mean}. However in these works it was mainly studied at equilibrium. In this case the ideal glass state, cooled down further undegoes a further replica symmetry breaking. This is different from the construction we have reviewed above, where instead we are looking at perturbed out-of-equilibrium glassy states. In this case one realizes the Gardner transition out-of-equilibrium and, possibly, well above the Kauzmann point. 

In the context of structural glass models, the importance of the Gardner transition emerged in \cite{kurchan2013exact}. Albeit in a slightly different theoretical framework with respect to the previous section, in this work it was understood  that infinite dimensional hard spheres, at sufficiently high pressure, undergo a Gardner transition. It was proposed that this was the essential ingredient to describe the criticality of the jamming transition and the anomalous low temperature properties of glasses. Beyond the Gardner point hard spheres are described by continuous replica symmetry breaking \cite{charbonneau2014fractal,charbonneau2014exact,rainone2016following}. What was a stable glass has become a metabasin of a hierarchical structure of states. This structure can be described in terms of the probability distribution of the the mean square displacement between different typical configurations belonging to the same glass metabasin. This probability distribution can be computed in the infinite dimensional limit. 

The formalism needed to compute the properties of the Gardner phase is close to the one found several years before in the context of spin glasses. In a nutshell one has a set of partial differential equations (a non-linear equation of the same form as the one discovered by G. Parisi in his solution of the Sherrington-Kirkpatrick model \cite{parisi1980sequence}, and another one with a Fokker-Plank structure analogous to what was found in spin glasses \cite{sommers1984distribution}).
With the addition of a closing equation, this formalism gives access to the cumulative distribution function of the mean square displacement $x(\Delta)$, see \cite{parisi2020theory} for more details.

The appearance of the Gardner transition, as described within the infinite dimensional solution, can be detected looking at a series of observables.
On approaching the Gardner point from the stable glass phase, one can look at the relaxation dynamics (Newtonian dynamics or Langevin dynamics) starting from a typical configuration of the glass. While in the stable glass phase the relaxation time is finite, at the Gardner point it diverges. One can focus on the dynamical mean square displacement defined as
\begin{equation}
\Delta(t) = \frac 1N \sum_{i=1}^N |\underline x_i(t) - \underline x_i(0)|^2.
\end{equation}
If $\{\underline x_i(0)\}$ denotes a typical configuration of the stable glass at initial time and $\{\underline x_i(t)\}$ the resulting configuration at time $t$, then $\Delta(t)$ relaxes to $\Delta_{EA}$ in an exponential way. Approaching the transition point, one finds a divergence of the relaxation time $\tau \sim |\phi-\phi_G|^{-\gamma}$ characterized by a non-universal critical exponent $\gamma$. Sitting exactly at the Gardner point the exponential relaxation is replaced by an algebraic one, $\Delta_{EA}-\Delta(t) \sim  t^{-a}$ being $a$ a non-universal dynamical critical exponent \cite{rainone2016following, parisi2020theory}.

The transition point can be also characterized by looking at diverging susceptibilities. The simplest one is the $\chi_4$ susceptibility defined as 
$\chi_4 = \lim_{t\to\infty} N(\overline{\Delta(t)^2)}-\overline{\Delta(t)}^2)$ and the overline stands for the collective  average over the realization of the initial conditions and thermal noise. Approaching the Gardner point one has that $\chi_4\sim 1/|\phi-\phi_G|$. Other diverging susceptibilities can be defined from the sample-to-sample fluctuations of non-linear elastic moduli \cite{biroli2016breakdown}. They imply a breakdown of elastic behavior replaced by  plastic responses upon strain deformation. Therefore while in the stable glass phase the response of the glass is elastic and reversible, in the Gardner phase this behavior is replaced by anomalously large responses to small perturbations, typically manifested through avalanche-like phenomenology. Therefore the Gardner phase is said to be marginally stable.
Finally, crossing the transition point, zero field cooled and field cooled shear responses start to differ \cite{jin2017exploring} and in the Gardner phase relaxation dynamics is expected to exhibit aging behavior.

It is useful to provide a short overview of which systems, in the infinite dimensional limit, have been shown to have a Gardner phase and under which circumstances.

Hard spheres under compression undergo always the Gardner transition. Therefore at sufficiently high pressures, any hard sphere glass, including the ideal one, is a Gardner glass \cite{kurchan2013exact, rainone2015following}.
The transition happens sooner the closer the parent glass is to the dynamical glass transition point. 
The same happens in soft spheres, like harmonic spheres \cite{biroli2018liu} or spheres interacting through a WCA potential \cite{scalliet2019marginally}. Remarkably, as was found in the latter case, there are situations and interaction potentials for which one can observe a Gardner transition only if the parent glass is sufficiently unstable and close to the mode-coupling point.
Harmonic spheres are interesting since they interpolate between the low density hard spheres regime and the high density soft sphere one. At zero temperature and as a function of the density, these two regimes are separated by the jamming transition \cite{liu2010jamming}. This can be seen as the point where zero temperature soft spheres trapped in a glass, cannot be arranged without making them overlap if further compressed. When $d\to \infty$, it is found that the jamming transition is always surrounded by a dome of a marginally stable Gardner phase \cite{biroli2016breakdown, scalliet2019marginally}.
Finally the effect of shear strain and combined strain deformation and compression can be also studied \cite{rainone2015following, rainone2016following, urbani2017shear,biroli2018liu,altieri2019mean}. Also in this case, if the glass is at sufficiently high pressure and low temperature, once strained it undergoes a Gardner transition.

\section{Sitting at the bottom of the landscape: marginal stability and scale invariance}
In the previous section we focused on the Gardner transition and on its characterization.
In this section we would like to give a short overview of what are the consequences for extremely low temperature glasses and high pressures granular glasses.

In the infinite pressure limit, hard spheres glasses jam and form rigid random configurations. 
The outcome of the infinite dimensional analysis, in the jamming limit, is that \cite{charbonneau2014fractal, charbonneau2014exact}:
\begin{itemize}
\item the Gardner phase, at infinite pressure, gives rise to isostaticity which is the property such that the number of contacts between spheres equals the number of degrees of freedom. This property emerges from first principles and it is a direct consequence of the marginal stability of the replica symmetry broken solution (full replica symmetry breaking).
\item at jamming one can compute the distribution of contact forces $f$ as well as small gaps $h$ between almost touching spheres. Both distributions are controlled by two critical exponents. At small argument they repsectively behave as $P(f)\sim f^{\theta}$ and $P(h)\sim h^{-\gamma}$ with $\theta=0.42311\ldots$ and $\gamma=0.41269\ldots$.
\item On approaching the jamming point from the hard sphere side one can show that the cage size $\Delta_{EA}$ (or Debye-Waller factor) decreases to zero with a power law of the pressure $p$. The critical exponent can be computed as well and one gets that $\Delta_{EA}\sim p^{-\kappa}$ with $\kappa=1.41574\ldots$.
\item on approaching the jamming point from the jammed phase of harmonic soft spheres, the vibrational density of states displays a plateau at low frequencies and this has been shown using a simplified model for infinite dimensional harmonic spheres \cite{franz2015universal}.
\end{itemize}

All these properties were shown to appear in extensive numerical simulations \cite{o2003jamming, liu2010jamming,wyart2005effects,brito2009geometric,lerner2013low,charbonneau2012universal}
Remarkably, the values of the critical exponents compare well with numerical simualtions in finite dimensions, see \cite{charbonneau2021finite} for the most recent investigation. 

The critical exponents, (as well as isostaticity) arise in a quite remarkable way from the formalism describing the Gardner phase. In the jamming limit, the equations describing the ultrametric structure of states within the glass metabasin, displays a scaling regime controlled by a set of critical exponents which imply the ones of forces, gaps and cage size. These exponents can be fully determined by solving numerically the corresponding equations.

This scaling regime can be generalized even beyond jamming. Recently, spheres interacting with a linear ramp potential (and therefore called linear spheres) have been shown to give rise to a full, jamming critical phase \cite{franz2020critical}. Above jamming if one considers linear spheres at zero temperature, one finds that they sit in local minima of the potential energy landscape which are isostatic (with a more subtle definition of isostaticity) and described by a set of power laws controlling the microstructural properties of these configuration. Again, these power laws can be found within mean field theory \cite{franz2019critical} and they emerge from a more complicated scaling solution of the equations describing the corresponding Gardner phase. Remarkably, non-trivial algebraic identities imply that the critical exponents arising in this case coincide with the ones at jamming point.

Both at jamming \cite{franz2017mean} and in mean field models of jammed linear spheres \cite{franz2021surfing}, the critical exponents computed from the infinite dimensional solution can be shown to control the avalanche size distribution of perturbed packings. Finally, the very same critical exponents are found to describe the power law divergence of the shear modulus of hard sphere glasses at very high pressure close to jamming \cite{yoshino2014shear}.

Away from jamming, in the jammed phase of harmonic soft spheres it was shown through an equivalent simpler model \cite{franz2015universal} that the Garnder phase can give rise to a vibrational density of states $D(\omega)\sim \omega^2$ populated by delocalized low frequency eigenvectors.

Recently this picture has been challenged by numerical simulations showning that soft glasses sit in local minima which are typically gapless but with a spectrum better described by $D(\omega)\sim \omega^4$ and populated by quisi-localized eigenvectors, see \cite{lerner2021low} and references therein for a recent review.
Therefore in \cite{bouchbinder2021low, rainone2021mean, folena2022marginal} a spin glass model having a schematic Gardner phase has been proposed showing that the Gardner phase may give rise to pseudogapped spectrum of quasi-localized modes following an $\omega^4$ law, see \cite{folena2022marginal} for more details.  
One of the main ingredient of the model is the presence of local heterogeneities which are shown to drive the form of the soft tail of the vibrational density of states. 
How to incorporate these effects in simple models of structural glasses in infinite dimensions remains to be done.

\section{Beyond mean field theory}
Up to now we have been describing the emergence of the Gardner transition and its consequences within the mean field limit of infinite spatial dimensions.
Here we will give a very short review on what is known beyond mean field theory.

From a critical phenomena point of view, the Gardner transition is in the same universality class as the spin glass transition in a field \cite{urbani2015gardner} since at this point only replica symmetry breaks down. To go beyond mean field theory one can perform perturbative renormalization group computations at criticality \cite{urbani2015gardner} which show that, as for the spin glass problem  \cite{bray1980renormalisation}, the Gaussian fixed point of the RG equations looses stability below the upper critical dimension and no other stable fixed point is found. Two-loops renormalization group computations show that a new fixed point can be found below the upper critical dimension but in a non-perturbative region of coupling constants \cite{charbonneau2017nontrivial}. However, the nature of the low temperature phase cannot be accessed by analyzing the theory at criticality. 

It was proposed in \cite{parisi2012replica} that if the spin glass transition in a field arises, it could be controlled by a critical point at zero temperature. 
Recently it has been shown \cite{folena2022marginal} that, within mean field theory, this transition can be driven by two different mechanisms. On the one hand one can have the appearance of a fat tail ($D(\omega)\sim \omega^2$) of low frequency delocalized eigenmodes in the density of states giving rise a divergence of the spin glass susceptibility. This scenario may not be robust in finite dimensions since getting delocalized excitations at the edge of the spectrum is expected to be hard unless some symmetries are present at long wavelengths (for example translational symmetry giving rise to phonons). On the other hand, in \cite{folena2022marginal} it was shown that one can have a transition driven by the appearance of a finite density of non-linear excitations whose nature is captured by the replica symmetry breaking solution. A field theory analysis focused on this second scenario has been developed recently to go beyond mean field theory \cite{urbani2022field} and it deserves further investigation.

% for BibTeX users
%\bibliographystyle{unsrt}
%\bibliography{refs}      % pls. call your database here

\end{document}